\pdfoutput=1
\documentclass[acmlarge,screen,nonacm]{acmart}

\AtBeginDocument{%
  }

\usepackage{subcaption}
\usepackage{xspace}
\usepackage{xurl}
\usepackage{listings}
\lstdefinestyle{prompt}{
  basicstyle=\footnotesize\ttfamily, columns=fullflexible, keepspaces=true,
  breaklines=true, breakatwhitespace=true, breakindent=1.5em,
  frame=single, framerule=0.4pt, rulecolor=\color{black!25},
  backgroundcolor=\color{black!4}, framesep=0.6em,
  xleftmargin=0.6em, xrightmargin=0.6em, aboveskip=0.4em, belowskip=0.9em,
  showstringspaces=false}
\newcommand{\model}[1]{\nolinkurl{#1}}

\makeatletter
\let\hyxmp@parse@acmart\relax
\makeatother
\begin{document}

\title[Unpacking Aging Representations and Stereotypes in LLM-generated Stories]{“Despite her age, Margaret was eager to make a change”: Unpacking Aging Representations and Stereotypes in LLM-generated Stories}

\author{MD. Roqunuzzaman Sojib}
\authornote{These authors contributed equally to this research.}
\email{sojib@cse.buet.ac.bd}
\affiliation{%
  \department{Department of Computer Science and Engineering}
  \institution{Bangladesh University of Engineering and Technology}
  \city{Dhaka}
  \country{Bangladesh}}
\author{Mohammad Sadat Hossain}
\authornotemark[1]
\email{sadat@cse.buet.ac.bd}
\affiliation{%
  \department{Department of Computer Science and Engineering}
  \institution{Bangladesh University of Engineering and Technology}
  \city{Dhaka}
  \country{Bangladesh}}
\author{Nafis Tahmid}
\authornotemark[1]
\email{ntahmid@cse.buet.ac.bd}
\affiliation{%
  \department{Department of Computer Science and Engineering}
  \institution{Bangladesh University of Engineering and Technology}
  \city{Dhaka}
  \country{Bangladesh}}
\author{Shattik Islam Rhythm}
\authornotemark[1]
\email{shattik@cse.buet.ac.bd}
\affiliation{%
  \department{Department of Computer Science and Engineering}
  \institution{Bangladesh University of Engineering and Technology}
  \city{Dhaka}
  \country{Bangladesh}}
\author{Novia Nurain}
\email{nnurain@cse.buet.ac.bd}
\affiliation{%
  \department{Department of Computer Science and Engineering}
  \institution{Bangladesh University of Engineering and Technology}
  \city{Dhaka}
  \country{Bangladesh}}

\renewcommand{\shortauthors}{Sojib, Hossain, Tahmid, Rhythm, and Nurain}

\authorsaddresses{Authors' Contact Information:
  MD. Roqunuzzaman Sojib, \href{mailto:sojib@cse.buet.ac.bd}{sojib@cse.buet.ac.bd};
  Mohammad Sadat Hossain, \href{mailto:sadat@cse.buet.ac.bd}{sadat@cse.buet.ac.bd};
  Nafis Tahmid, \href{mailto:ntahmid@cse.buet.ac.bd}{ntahmid@cse.buet.ac.bd};
  Shattik Islam Rhythm, \href{mailto:shattik@cse.buet.ac.bd}{shattik@cse.buet.ac.bd};
  Novia Nurain, \href{mailto:nnurain@cse.buet.ac.bd}{nnurain@cse.buet.ac.bd}.
  Department of Computer Science and Engineering, Bangladesh University of
  Engineering and Technology, Dhaka, Bangladesh}

\begin{abstract}
As large language models (LLMs) increasingly generate everyday stories, their portrayals can shape expectations about aging. Prior work has largely examined age stereotypes in short responses, leaving unanswered whose lives models default to imagining and how age shapes characters’ roles in complete narratives. We analyze 594 stories generated by Gemini, GPT, and Llama using text embeddings, LLM-assisted coding, and qualitative analysis. When age was unspecified, stories typically featured younger adults, with middle-aged protagonists largely absent. Older adults were often portrayed through loss, familiar routines, and help from younger characters. Even positive stories sometimes treated older adults’ achievements as surprising or portrayed others making decisions on their behalf. We suggest evaluating not only the tone of generated stories, but also who appears, what opportunities they receive, and who makes decisions. Involving older adults in these evaluations can guide writing tools toward more varied, less stereotyped portrayals of later life.
\end{abstract}



\maketitle

\section{Introduction}
\label{sec:intro}

Stories are one of the main ways people learn what a life is supposed to look like, and a growing share of them are now written by large language models (LLMs). People use LLMs to draft fiction~\cite{chakrabarty2024creativity}, to prepare reading and teaching material, and to generate bedtime stories for children~\cite{rooein2025biased}. A request such as ``write a story about saving money'' names only a topic. The model decides the rest: who the protagonist is, how old they are, what their days look like, what goes wrong, and who sets it right. When these choices repeat across millions of requests, they become a portrayal of whole groups of people. For older adults, the stakes are high. Ageism is widespread, with an estimated one in two people worldwide holding ageist attitudes toward older adults ~\cite{who2021global}, and it is linked to worse health outcomes~\cite{chang2020global}. Age stereotypes internalized in childhood can become self-relevant as individuals grow older, shaping their perceptions of aging ~\cite{levy2009stereotype}. Stories that narrow later life, therefore, affect both how younger readers imagine old age and how older readers are taught to live it.

Despite these stakes, age has received far less attention than gender or race in AI fairness research~\cite{stypinska2023ai,chu2022digital}. Stypinska describes \emph{AI ageism} as a set of harms spanning biased data and algorithms, age stereotypes, and the exclusion of older adults from AI discourse and development~\cite{stypinska2023ai}. Gallistl et al.\ further highlight the underrepresentation of older adults in AI policy and the limited protections against algorithmic ageism~\cite{gallistl2025aiaging}. Existing studies of ageism in generative AI have identified stereotyped portrayals of older adults in text-to-image generation~\cite{allen2023age}, while research on language technologies has examined sentiment differences between age-related terms~\cite{diaz2018addressing}, stereotypes in sentence completions~\cite{kamruzzaman2024investigating} and short question-answering tasks~\cite{parrish2022bbq}. However, these approaches offer limited insight into how stereotypes shape characters, relationships, and narrative outcomes. This is particularly important for understanding benevolent ageism, in which older adults are treated warmly but assumed to be less capable and in need of help~\cite{cuddy2005old,cary2017ambivalent}, a pattern also observed in LLMs at levels comparable to those measured in humans~\cite{chen2025ai}. Complete stories provide an opportunity to examine how both hostile and benevolent ageism emerge through who appears, what challenges protagonists face, and who ultimately resolves their problems. Yet age-related stereotyping in LLM-generated narratives remains underexplored. Hence, in this work, we seek to answer the following question: \textbf{How do LLM-generated stories portray older adults?}

To answer this question, we analyzed 594 stories generated by Gemini, GPT-5, and Llama across 22 everyday \emph{life factors}. For each factor, we generated stories about younger adults, older adults, and stories without specifying an age. Using text embeddings, LLM-assisted coding, and human qualitative analysis, we found that when age was unspecified, the models predominantly portrayed younger adults, while middle-aged and older adults were rarely represented. Stories about older adults often centered on loss, past experiences, and maintaining independence, whereas younger adults were portrayed as forming new relationships and exploring opportunities. Older adults were also frequently shown receiving help from younger people, who sometimes made decisions on their behalf. Besides, the stories often portrayed older adults' growth, confidence, and achievements as remarkable because of their age.

Building on our findings, the work makes the following contributions: \textbf{First}, we provide an empirical understanding of how LLMs represent aging in everyday stories, examining implicit age defaults and differences in the experiences of older and younger protagonists. \textbf{Second}, we expand understandings of age-related representational bias by showing how younger adulthood becomes the assumed norm, leaving middle adulthood largely invisible, and how ageist assumptions can persist even within positive narratives through the framing of assistance, capabilities, and achievements. \textbf{Third}, we highlight methodological implications for evaluating aging representations and assumptions by combining diverse approaches, such as story similarity, LLM-assisted coding, and qualitative analysis, to examine both who is represented and how narratives are constructed. \textbf{Finally}, we offer design implications for identifying recurring age-related assumptions, supporting more diverse portrayals of later life, and involving older adults in evaluating these representations.

\section{Background}
\label{sec:related}

\subsection{Ageism and representations of later life}
\label{sec:rw-ageism}

Ageism research has identified several recurring stereotypes about older adults, who are often portrayed as physically or cognitively impaired, less competent, lonely, or socially withdrawn. Other stereotypes depict them as rigid or out of touch with change~\cite{kite2005attitudes,cuddy2005old,hummert1994stereotypes}. Hummert et al.\ identified distinct images of older adults, including negative types such as the \emph{severely impaired}, the \emph{despondent}, and the \emph{recluse}, alongside positive types such as the \emph{golden ager} and the \emph{perfect grandparent}~\cite{hummert1994stereotypes}. These representations can reflect two broad forms of ageism: hostile and benevolent~\cite{cary2017ambivalent}. Hostile ageism involves resentment, exclusion, or discrimination, often portraying older adults as burdens or as occupying resources that should pass to younger generations. North and Fiske further describe expectations that older people should yield resources and opportunities to younger generations and avoid behaviours considered inappropriate for their age~\cite{north2012inconvenienced,north2013act}. Such attitudes can reinforce discrimination and barriers to participation~\cite{nelson2002ageism,swift2017risks}. In contrast, benevolent ageism expresses assumptions of incapacity through warmth and protection. The stereotype content model characterizes older adults as warm but low in competence, inviting pity and paternalism rather than contempt~\cite{fiske2002model,cuddy2002doddering,cuddy2005old}. This can manifest as unsolicited assistance or decisions made on older adults' behalf~\cite{cary2017ambivalent}. Even seemingly positive generalizations, such as assuming all older adults are wise, can be reductive, reflecting what Palmore termed positive ageism~\cite{palmore1999ageism}. These forms can coexist, portraying older adults as deserving care while simultaneously limiting their agency.

Such stereotypes extend beyond interpersonal attitudes into media and technology. Research on advertising and visual media has found that older adults are often underrepresented or depicted through a narrow range of stereotypes~\cite{eisend2022older,loos2022visual}. An analysis of two centuries of American English similarly found that portrayals of older people became increasingly negative over time~\cite{ng2015increasing}. These patterns are particularly relevant to LLMs, which learn from large collections of human-produced text and may reproduce biases embedded in their training data~\cite{bender2021dangers}. In technology design, older adults have frequently been framed in terms of decline, dependency, and care, positioning aging as a problem for technology to address~\cite{vines2015ageold}. More recent work calls for greater recognition of older adults' agency and diverse experiences~\cite{lazar2025hci}. For instance, choosing not to use a technology may reflect a deliberate decision rather than an inability to use it~\cite{knowles2018wisdom}, while older bloggers demonstrate how digital platforms can support self-expression and social participation~\cite{brewer2016tell}. Nevertheless, ageist assumptions continue to shape understandings of older adults' relationships with technology~\cite{mannheim2023ageism}, and digital platforms and AI systems can reproduce existing inequalities and age-related biases~\cite{rosales2020ageism,chu2022digital}.

These representations matter because age stereotypes can be internalized across the life course and subsequently shape how people perceive and treat themselves as they age, as described by stereotype embodiment theory~\cite{levy2009stereotype}. Moreover, age stereotypes vary across life domains, including work, family, health, and finances~\cite{kornadt2011contexts}. Building on these perspectives, we examine portrayals across 22 everyday life factors, considering both overtly negative representations of decline, isolation, and incompetence and seemingly positive portrayals that reinforce dependency or constrain agency. This raises an important question for generative AI: when LLMs create stories about older adults, do they reproduce familiar narratives of decline and dependency, or represent a broader range of experiences, choices, and identities?

\subsection{LLMs as storytellers and auditing representational bias}
\label{sec:rw-storytellers}

Storytelling has become an increasingly common creative application of LLMs, ranging from fiction drafting and revision~\cite{yuan2022wordcraft,chakrabarty2024creativity} to educational materials~\cite{xiao-etal-2023-evaluating} and personalized bedtime stories~\cite{rooein2025biased}. When given only a topic and desired length, models must fill in unspecified details, including protagonists, their circumstances, the problems they face, and how these are resolved. These choices make generated stories useful for studying implicit model assumptions, particularly about who is considered ordinary and what roles different people occupy. For example, LLM-generated descriptions of marginalized groups contain more racial stereotypes than comparable human-written descriptions~\cite{cheng2023marked}, while stories without specified identities often omit minoritized groups or place them in subordinate roles~\cite{shieh2026intersectional}. Similar representational patterns have been identified for nationality~\cite{nguyen2026representational} and gender assignments to animal characters~\cite{finkley2026neutrality}. Thus, representational bias can influence both who appears in generated narratives and the roles they assume.

Research on auditing language bias has evolved from examining word associations~\cite{caliskan2017semantics} to evaluating generated text through measures such as sentiment, toxicity, and regard~\cite{sheng2019woman,dhamala2021bold}. However, these measures cannot capture all representational harms, which may also arise from a text's underlying assumptions about a person, even when its language is neutral or positive~\cite{blodgett2020language}. Studies of longer generated texts illustrate that LLM-generated stories have associated female characters more closely with family and appearance while assigning them less power than male characters~\cite{lucy2021gender}; generated reference letters have also exhibited gender differences in how individuals are described~\cite{wan2023kelly}. Similarly, studies of disability have identified portrayals of disabled people as dependent on others, including the stereotype of an able-bodied savior~\cite{gadiraju2023wouldn}, and deficit-based assumptions embedded in otherwise inclusive descriptions of autism~\cite{park2025autistic}. These findings demonstrate that representational bias can emerge through the roles, abilities, and agency assigned to characters, rather than through explicitly negative language alone.

Research on age bias in language models has primarily examined word associations, sentiment, and question answering~\cite{diaz2018addressing,parrish2022bbq,kamruzzaman2024investigating}. More recent studies have investigated age-related stereotypes in LLM-generated responses, including assumptions about older adults' technology use, physical and cognitive decline, and occupational roles~\cite{dewan2025examining}. Chen et al.\ found that LLMs expressed less hostile but comparable levels of benevolent ageism relative to humans, with AI-attributed ageist statements potentially having a stronger influence on people's attitudes~\cite{chen2025ai}. Age-related biases have also been identified in AI-generated images~\cite{kamelski2024generative} and in the intersection of age and gender in LLM-generated content~\cite{guilbeault2025age}.

Despite these advances, less attention has been paid to how age stereotypes shape complete fictional narratives. Examining stories requires looking beyond individual descriptions: an older character may be portrayed positively while still being denied agency, made dependent on others, or unable to resolve their own problems. Ageism can therefore emerge through how narratives distribute agency, assistance, and problem-solving, even without explicitly negative language. In particular, little is known about the ages LLMs implicitly assign to protagonists when age is unspecified, how younger and older characters are portrayed across everyday situations, and how their narrative agency and dependence differ. We address these gaps by examining both hostile and benevolent ageism in LLM-generated stories, focusing on character portrayals, narrative roles, and the distribution of agency in resolving central conflicts.

\section{Method}
\label{sec:method}

\begin{figure*}[!t]
  \centering
  \includegraphics[width=\linewidth]{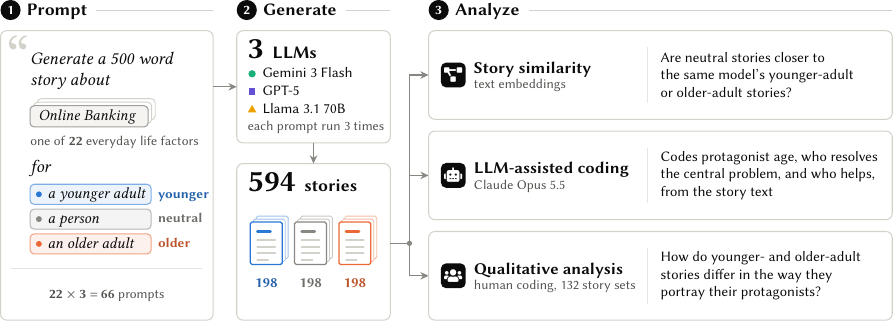}
  \caption{Study design. Each prompt combined one of 22 everyday life factors with one of three age conditions, which differ only in the age phrase. Three LLMs answered each of the 66 prompts three times, yielding 594 stories (198 per condition). We analyzed the stories in three ways: text-embedding similarity between each model's neutral and age-specified stories, LLM-assisted coding of each story's protagonist, and human qualitative analysis of 132 younger- and older-adult story sets.}
  \Description{A left-to-right flow diagram in three stages. Stage 1, Prompt: the template ``Generate a 500 word story about {factor} for ...'', with the factor slot shown as a stack of cards (one of 22 everyday life factors, e.g., Online Banking) and the age slot shown as three options: ``a younger adult'' (younger), ``a person'' (neutral), and ``an older adult'' (older); 22 times 3 gives 66 prompts. Stage 2, Generate: three LLMs (Gemini 3 Flash, GPT-5, Llama 3.1 70B), each prompt run three times, produce 594 stories, drawn as three stacks of 198 stories, one per condition. Stage 3, Analyze: arrows lead from the stories to three analyses. Story similarity with text embeddings asks whether neutral stories are closer to the same model's younger-adult or older-adult stories. LLM-assisted coding with Claude Opus 5.5 codes protagonist age, who resolves the central problem, and who helps, from the story text. Qualitative analysis, human coding of 132 story sets, asks how younger- and older-adult stories differ in the way they portray their protagonists.}
  \label{fig:method}
  \vspace{-0.5cm}
\end{figure*}

To better understand how various LLMs represent aging and older adults, we analyzed stories generated by three LLMs across different aspects of everyday life. We deliberately used minimal prompting under three age conditions (e.g., older adults, younger adults, and age-neutral) to explore how LLMs construct age-related representations through their choices of characters, settings, experiences, and narratives, drawing on their existing learned patterns without additional empirical information or prescribed narrative details. By analyzing the differences across the three conditions, we sought to understand which age groups LLMs portray when age is unspecified and how specifying younger or older adulthood shapes the protagonist's life circumstances, opportunities, challenges, and social roles. We also investigated whether these portrayals reflect any recurring societal assumptions or stereotypes about aging. We used a mixed-methods approach to capture both broader patterns and nuanced portrayals of aging in LLM-generated stories. We conducted a quantitative analysis to identify patterns in age representation and narrative roles. Additionally, we conducted qualitative analysis to understand how LLMs construct portrayals of aging through recurring narratives, assumptions, and stereotypes. It is to be noted that our goal was to identify recurring patterns of age representation across the generated stories rather than systematically benchmark differences in representational bias between individual models. Fig. \ref{fig:method} illustrates an overview of our study design and analysis. 

\subsection{Generating Stories}
\label{sec:generation}

To generate stories, the research team curated a list of five domains of everyday life informed by prior research on aging \cite{walsh2017social} and demographic biases \cite{wang2024newjob}. Walsh et al. \cite{walsh2017social} identified several dimensions of social exclusion in later life and highlighted how older adults' opportunities and participation may be shaped by their financial resources, social relationships, access to services, and civic engagement. Building on this work, we identified contexts in which LLMs might reproduce assumptions about older adults' capabilities, independence, and participation in everyday life. Additionally, Wang et al. \cite{wang2024newjob} demonstrated how demographic biases in image generation models can emerge through representations of ordinary professions, activities, and objects, even when prompts do not explicitly specify demographic characteristics. This informed our selection of diverse everyday factors to investigate how LLMs associate different experiences, roles, and activities with age. Drawing on these works, we identified five domains encompassing 22 factors: (1) \textit{economic life} (employment, personal finances, income stability, rising prices, future savings, and debt repayment), (2) \textit{careers} (professional development, new skills, and career growth), (3) \textit{social life} (community events, social belonging, professional networking, making friends, and social status), (4) \textit{civic life} (civic duty and current politics), and (5) \textit{technology} (new technology, transportation, online banking, automated systems, online privacy, and social media). For generating each story, we developed a generic prompt template: ``\texttt{Generate a 500 word story about [factor] for [age condition]}.'' We used three age conditions: (1) older adult, (2) younger adult, and (3) age-neutral, represented in the prompts as an older adult, a younger adult, and a person, respectively. For example, for older adults the prompt would state: \texttt{``Generate a 500 word story about online banking for an older adult}.'' We specified 500 words to allow sufficient development of characters, circumstances, and narrative outcomes while limiting unnecessary elaboration. We generated stories using three popular and publicly-available models: (1) Gemini 3 Flash (\texttt{google/gemini-3-flash-preview}) \cite{googledeepmind2025gemini3flash}, (2) GPT-5 (\texttt{openai/gpt-5}) \cite{openai2025gpt5}, and (3) Llama 3.1 70B Instruct (\texttt{meta-llama/llama-3.1-70b-instruct}) \cite{grattafiori2024llama}. To account for variability in LLM-generated responses, we submitted each factor and age-condition prompt three times independently to each model, resulting in 594 stories (22 factors $\times$ 3 conditions $\times$ 3 models $\times$ 3 generations), with a median length of 543 words.

\subsection{Data Analysis}
\label{sec:analysis}
We used a mixed-methods approach to analyze the generated stories. The quantitative analysis focused on broader patterns in age representation and narrative roles, while the qualitative analysis explored how LLMs constructed portrayals of aging through recurring narratives, assumptions, and stereotypes.

\subsubsection{Quantitative  Analysis}
We conducted quantitative analysis using three complementary approaches to investigate how LLMs represent age across the generated stories. We first used text embeddings to assess whether age-neutral stories were more similar to stories about younger or older adults, helping us understand which age groups LLMs tend to represent when age is unspecified. However, understanding age representation also requires investigating how LLMs characterize different age groups, particularly their perceived capabilities, agency, and dependence on others, which story similarity alone cannot reveal. To address this limitation, we used another LLM to systematically code the generated stories, identifying protagonists' ages, who resolved their challenges, and who provided them with assistance. Finally, we applied statistical analyses to assess the consistency of the observed patterns and differences across models and everyday life factors, while accounting for variation among repeated story generations.

\textit{\underline {Story Similarity}}: To assess which age group LLMs tend to represent when age is unspecified, we compared age-neutral stories with their younger- and older-adult counterparts using gemini-embedding-2-preview \cite{shanbhogue2026geminiembedding2native}, a text embedding model that converts text into numerical vectors to measure semantic similarity. For each model $m$ and factor $f$, we normalized the story embeddings to unit length and separately averaged the embeddings of the three independently generated younger-adult stories ($\bar{y}_{m,f}$) and older-adult stories ($\bar{o}_{m,f}$) to obtain a representative embedding for each age condition. We then calculated the default-age lean of each age-neutral story $i$, represented by embedding $n_i$, as follows:

\begin{equation}
  \mathrm{lean}_i = \cos(n_i, \bar{y}_{m,f}) - \cos(n_i, \bar{o}_{m,f}),
\end{equation}

Here, $\cos$ denotes cosine similarity. Positive values indicate greater similarity to younger-adult stories and negative values indicate greater similarity to older-adult stories. We averaged these values across the three age-neutral generations to obtain a representative default-age lean score for each model and factor, accounting for variability in the stories.

\textit{\underline {LLM-Assisted Story Coding}}: Prior research has shown that LLMs can support structured content analysis \cite{gilardi2023chatgpt,tornberg2025large,xiao2023supporting}. Building on this work, we used Claude Opus 5.5 (claude-opus-5-5) \cite{anthropic2026opus55} to systematically code all 594 stories for protagonists' ages, problem resolution, and assistance received. Each story was processed in a separate Claude Code session, with tools and user settings disabled, using structured coding instructions without revealing the generating model or age condition. For protagonist age, we instructed the LLM to extract explicitly stated ages or estimate them from contextual cues, such as career stage, family relationships, and retirement, and provide a confidence rating. We used these ages to calculate a default-position score, indicating how closely the median age of neutral protagonists aligned with those in the younger- and older-adult conditions. To investigate protagonists' narrative roles, the coding prompt asked the LLM to identify who primarily resolved the central challenge (the protagonist, another character or institution, or both equally) and whether the protagonist received substantial assistance from someone younger, older, or approximately the same age. The complete coding instructions and calculation are provided in Appendix~\ref{app:prompts}.

\textit{\underline {Statistical Analysis}}: We applied statistical tests to assess the consistency of observed differences in story similarity, protagonist ages, problem resolution, and assistance across study conditions (i.e., model, age, factor). Because each prompt produced three stories, we averaged their scores for the default-age measures and kept the three stories together during resampling for problem resolution and assistance analyses. For story similarity, we used two-sided Wilcoxon signed-rank tests to assess whether average default-age lean scores differed from zero, both within each model across 22 factors and across models using factor-level averages. For protagonist ages, we averaged default-position scores across models for each factor and applied the same test against $0.5$ to assess whether age-neutral protagonists tended to be closer in age to younger or older protagonists. We calculated 95\% confidence intervals for both measures using 2,000 bootstrap resamples. For LLM-coded problem resolution and assistance, we used permutation tests to compare the proportions of protagonists resolving their own challenges, receiving cross-generational assistance, and receiving peer assistance between younger- and older-adult stories. We calculated 95\% bootstrap confidence intervals using 5,000 iterations, keeping repeated generations grouped during resampling. We considered differences significant when $p < 0.05$ and the confidence interval excluded zero.

\subsubsection{Qualitative Analysis}
We adopted thematic analysis \cite{braun2006using} to explore recurring aging representations, assumptions, and stereotypes in LLM-generated stories. We focused on 396 younger and older adult stories (198 per age condition) and excluded the 198 age-neutral stories to investigate how explicitly specifying age shaped the narratives. We conducted inductive thematic analysis in which all authors independently open-coded stories for a single factor (i.e., automated systems). For this factor, authors coded all three generations from each model under both age conditions. The researchers then compared their codes, discussed differences in interpretation, and resolved disagreements through research meetings to develop a shared codebook. We also incorporated deductive codes informed by prior research on age stereotypes and digital ageism \cite{levy2009stereotype, cuddy2005old} to capture established age-related assumptions. After developing the initial codebook through inductive and deductive coding, we assigned the remaining factors to researchers for further analysis. Our initial analysis across repeated generations revealed substantial similarities in narrative patterns. Therefore, we selected one story, generally the first generation, from each model and age condition for the remaining factors to reduce redundant coding while maintaining coverage across all factors and conditions. Throughout the analysis, the research team held asynchronous discussions and meetings to review and refine codes, organize them into categories, identify emerging patterns, and develop broader themes. We used the online qualitative analysis tool \textit{Taguette}\footnote{\url{https://app.taguette.org/}} to facilitate collaborative coding.

\section{Findings}
\label{sec:findings}

Our findings highlight how LLM-generated stories reflect assumptions about aging, both in the characters they portray and in how their lives, capabilities, and relationships are represented. We first investigate which age groups the models tend to portray when age is left unspecified. We then examine recurring patterns in how the stories portray older adults' social relationships, responses to changing circumstances, interactions with technology, and experiences of receiving assistance and protection, drawing comparisons with younger-adult narratives. Finally, we examine the contradictions within positive portrayals of aging, exploring how stories that celebrate older adults' capabilities and achievements may simultaneously reinforce assumptions and stereotypes about aging.

\subsection{Younger Adults Dominate Age-Neutral Stories}
\label{sec:f-default}

\begin{figure}[t]
  \centering
  \begin{subfigure}[t]{0.45\linewidth}
    \centering
    \includegraphics[height=6.2cm,width=\linewidth,keepaspectratio]{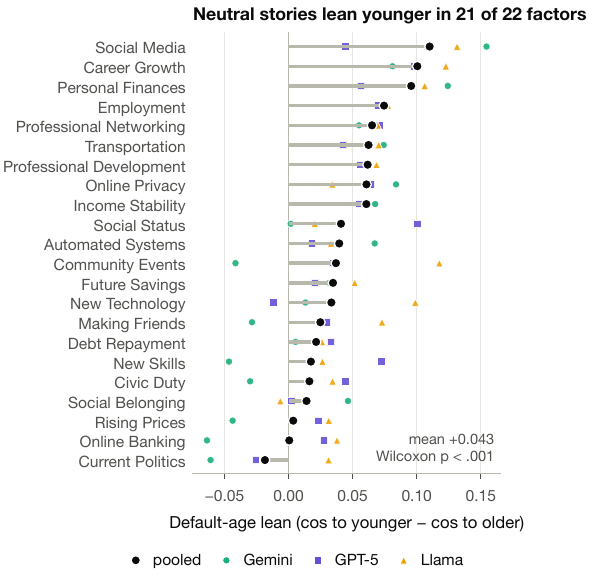}
    \caption{Default-age lean of neutral stories}
  \end{subfigure}\hfill
  \begin{subfigure}[t]{0.45\linewidth}
    \centering
    \includegraphics[height=6.2cm,width=\linewidth,keepaspectratio]{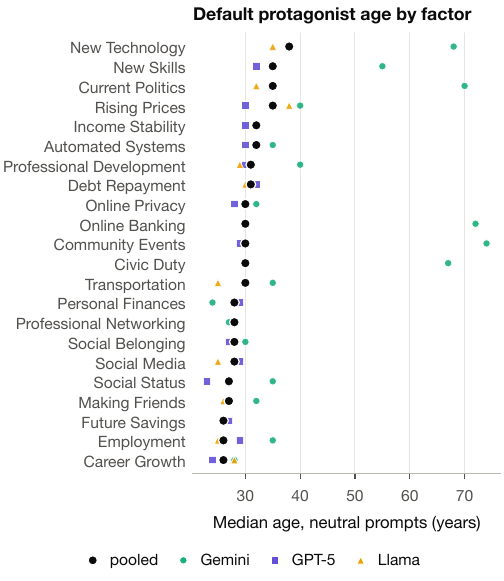}
    \caption{Median protagonist age in neutral stories}
  \end{subfigure}
  \caption{When no age is given, the models write about young adults. (a) Default-age lean for each factor: a neutral story's cosine similarity to the same model's younger-adult stories minus its similarity to that model's older-adult stories, averaged over the factor's neutral stories; positive values mean the neutral stories read younger. (b) Median LLM-coded age of the neutral protagonist. Black dots pool the three models; colored markers show Gemini (green circles), GPT-5 (purple squares), and Llama (yellow triangles).}
  \Description{Two dot plots with one row per life factor. In the left plot, 21 of 22 factors have a positive default-age lean, ranging from about 0.11 for social media down to about zero for online banking; only current politics is slightly negative. In the right plot, median neutral protagonist ages range from 26 for career growth, employment, and future savings to 38 for new technology; one model (Gemini) places the neutral protagonist near 70 for civic duty, community events, current politics, new technology, and online banking.}
  \label{fig:default}
  \vspace{-0.5cm}
\end{figure}

\textit{\underline{Neutral stories more closely resemble younger-adult stories}}: Across all three models, stories generated without an age specification were more similar to the same model's younger-adult stories than to its older-adult stories. As measured by default-age lean, this pattern appeared in 21 of the 22 factors after averaging across models (mean $+0.043$, $p < .001$; Figure~\ref{fig:default}a). The strongest lean toward younger adults appeared in social media, career growth, and personal finances. The difference was small for online banking and rising prices, while current politics was the only factor that leaned toward older adults. All three models showed a positive average lean, with the highest value for Llama ($+0.060$) and the lowest for Gemini ($+0.029$; Appendix~\ref{app:permodel}).

\begin{figure}[t]
  \centering
  \includegraphics[width=0.5\linewidth]{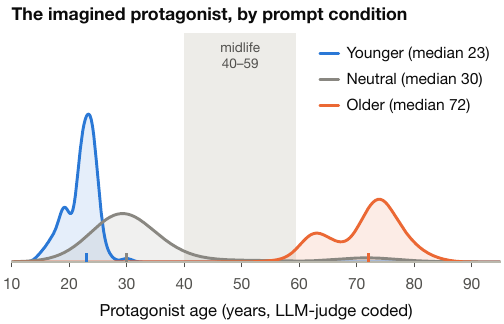}
  \caption{Distribution of LLM-coded protagonist age by prompt condition, pooled over models. The shaded band marks midlife (40--59). Ticks on the axis mark medians (rounded in the legend).}
  \Description{Three density curves. The younger-adult curve is a narrow peak around 23 years. The neutral curve is wider and peaks around 30 years with a long, thin right tail. The older-adult curve has peaks near 63 and 74 years. The shaded midlife band between 40 and 59 contains almost no density from any condition.}
  \label{fig:age}
  \vspace{-0.5cm}
\end{figure}

The ages assigned to protagonists further supported this tendency toward younger adulthood. Neutral stories had a median protagonist age of 30 years (IQR 27–34), compared with 23 years for younger-adult stories and 72 years for older-adult stories (Figure~\ref{fig:age}). The default-position measure placed neutral protagonists, on average, only 23\% of the way from the younger-adult median toward the older-adult median (95\% CI 17--28\%). The value was below the midpoint for all 22 factors ($p < .001$). Only 24\% of neutral stories explicitly stated a protagonist's age, so most ages were estimated from details such as career stage and family relationships. However, the median remained 30 for both high-confidence age estimates and stories that explicitly stated an age. Thus, the protagonist-age analysis reinforced the story-similarity findings, indicating that the models generally imagined younger adults when age was unspecified.

Despite this overall tendency, the ages assigned to neutral protagonists varied across factors and models. The median neutral protagonist age ranged from 26 years for career growth, employment, and future savings to 38 years for new technology (Figure~\ref{fig:default}b). Only 15 neutral protagonists were aged 60 or above. Most appeared in civic and technology factors, and all 15 came from Gemini. Thus, these older neutral protagonists did not reflect a pattern shared by the three models. Gemini also reused the name ``Arthur'' for 64 of its 66 older-adult protagonists and for 21 neutral protagonists, including some younger characters. Beyond the limited representation of older adults, midlife was also largely absent. We defined midlife as ages 40–59. Although age-neutral prompts allowed models to choose protagonists of any age, only 6\% fell within the 40–59 age range (Figure~\ref{fig:age}). The age-specific stories also tended toward widely separated age groups: all but two younger-adult protagonists were under 30, while most older-adult protagonists were at least 70. In these stories, ``younger'' often meant someone at the beginning of adult life and ``older'' someone in retirement. These findings show that age-neutral narratives predominantly centered younger adulthood, while middle adulthood remained largely unrepresented.

\subsection{Recurring Patterns in Portrayals of Later Life}
\label{sec:f-later-life}
Our findings reveal that LLM-generated stories often portray older adults' lives through loss, familiar routines, technological challenges, and reliance on younger people for assistance, reflecting common societal stereotypes about aging. In contrast, younger protagonists were more often depicted forming new relationships, pursuing opportunities, and navigating unfamiliar situations. These differences extended beyond the circumstances the characters encountered to the ways the stories framed their experiences, capabilities, opportunities, and relationships with others.

\label{sec:f-loss}
\textit{\underline {Social Life Framed by Loss, Memories, and Reconnection}}: LLMs often portrayed older adults' social lives through loss, memories, and the absence of loved ones, with social participation frequently framed as an attempt to rebuild diminished connections. For instance, one story described an older woman's circumstances: ``\textit{Her children had long since flown the coop, her husband had passed away, and her friends had either moved away or were no longer with her}'' (Llama, Social Belonging, Older). Another described a widower's home as ``\textit{less like a home and more like a vast, quiet museum}'' (Gemini, Community Events, Older). Even when family members appeared in older protagonists' lives, they were sometimes represented through memories, such as a character ``\textit{surrounded by photos of her children and grandchildren}'' (Llama, Career Growth, Older). Thus, these narratives often situated older adults' social experiences around remembering past relationships and recovering connections affected by loss.

In contrast, younger protagonists were more often portrayed through ongoing family interactions and opportunities to form new relationships. For example, a father asked his son over dinner, ``\textit{How's the compound interest treating you, son}?'' (Gemini, Future Savings, Younger). When the stories mentioned younger protagonists experiencing loneliness, it was often attributed to transitions or difficulties forming relationships rather than the loss of established ones. One younger character missed ``\textit{hallway hellos, lab partners}'' after moving away from familiar surroundings (GPT-5, Making Friends, Younger), while another was described as ``\textit{the shy and introverted type}'' who struggled to make friends (Llama, Social Belonging, Younger). Thus, while loneliness appeared across age groups, older adults' social lives were often framed around remembering and rebuilding lost connections, whereas younger adults' narratives emphasized maintaining existing relationships and establishing new ones.

\textit{\underline {Later Life Anchored in Familiarity and Past Experience}}: Older protagonists' stories often situated them within familiar environments and established routines, with changing circumstances narrated in relation to their existing ways of life rather than new beginnings. For example, Miriam maintained her financial records in a monthly ledger, following ``\textit{a habit her mother had taught her during the lean years}'' (GPT-5, Income Stability, Older). Similarly, Mariela encountered changing professional requirements despite her longstanding expertise: ``\textit{She had managed chaos before—drivers, storms, sudden shortages—but now she needed vocabulary and certificates to match the work her hands already knew}'' (GPT-5, Career Growth, Older).

While prior research identified resistance to change as a recurring stereotype of older adults in LLM-generated responses \cite{dewan2025examining}, our analysis revealed another dimension of how aging and change were portrayed. Older protagonists were not necessarily resistant to change. Rather, their accumulated knowledge and experience often became resources for navigating unfamiliar circumstances. For instance, when an automated logistics system failed, Arthur drew on his practical expertise to propose an alternative: ``\textit{The sensors are down, but the roads aren't}'' (Gemini, Career Growth, Older). In contrast, a younger protagonist in a similar logistics setting explored automation after discovering online tutorials, driven by ``\textit{a spark of curiosity}'' (Gemini, Career Growth, Younger). Thus, the narratives often grounded older adults' competence and responses to change in established practices and past experiences, while portraying younger adulthood through experimentation, developing capabilities, and future possibilities.

\textit{\underline{Technology Framed Around Maintaining Independence}}: In older-adult stories, technology was often introduced as a means of addressing everyday difficulties and maintaining independence, rather than exploring new possibilities. For instance, a granddaughter introduced a new device to her grandmother by explaining, ``\textit{They'll make things bigger or smaller, read out labels, remind you about your pills}'' (GPT-5, New Technology, Older). Such narratives frequently associated technology with accessibility, assistance, and managing everyday needs. Older protagonists were also sometimes portrayed as hesitant or unfamiliar with technology, even before encountering specific difficulties. One character approached a touchscreen with ``\textit{his pointer finger hovering cautiously over the glass as if a wrong move might trigger an alarm}'' (Gemini, Online Banking, Older). These portrayals often positioned technology as something older adults needed support to navigate.

In contrast, younger protagonists were more often portrayed as technologically competent, using technology to explore opportunities and develop new capabilities. For example, one younger analyst ``\textit{taught himself Python through online tutorials during his lunch breaks and built a script}'' to automate repetitive work (Gemini, Career Growth, Younger). Another was described as someone who could navigate complex software and manage spreadsheets ``\textit{in his sleep}'' (Gemini, New Skills, Younger). Thus, the distinction was not simply whether protagonists could use technology, but the purposes assigned to it. While technology often helped older adults address difficulties and maintain existing capabilities, younger protagonists more frequently encountered it as a means of experimentation, innovation, and expanding their opportunities.

\begin{figure}[!t]
  \centering
  \begin{subfigure}[t]{0.45\linewidth}
    \centering
    \includegraphics[height=3.4cm,width=\linewidth,keepaspectratio]{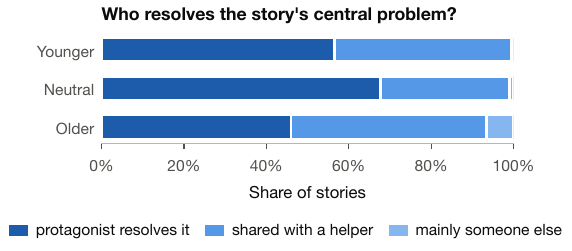}
    \caption{Who resolves the central problem}
  \end{subfigure}\hfill
  \begin{subfigure}[t]{0.45\linewidth}
    \centering
    \includegraphics[height=3.4cm,width=\linewidth,keepaspectratio]{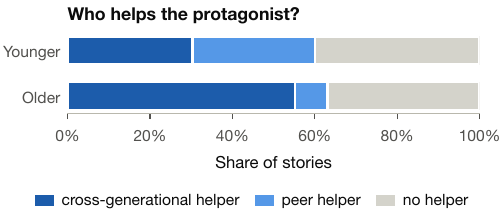}
    \caption{Who helps the protagonist}
  \end{subfigure}
  \caption{Problem resolution and help, pooled over models (LLM-coded). (a) Share of stories in which the protagonist resolves the central problem, shares the resolution equally with a helper, or relies mainly on someone else. A protagonist who resolves the central problem may still have been helped at some point in the story, so the two panels overlap. (b) Share of younger- and older-adult stories in which the protagonist is helped by someone from another generation (an older person helping a younger protagonist, or a younger person helping an older one), by a peer, or by no one; stories with a helper of unclear age are excluded.}
  \Description{Two stacked horizontal bar charts. In the left chart, the protagonist resolves the problem in about 57 percent of younger-adult stories, 68 percent of neutral stories, and 46 percent of older-adult stories; most of the rest are shared with a helper, and about 7 percent of older-adult stories rely mainly on someone else. In the right chart, a cross-generational helper appears in about 30 percent of younger-adult stories and 55 percent of older-adult stories, while a peer helper appears in about 30 percent of younger-adult stories and 8 percent of older-adult stories.}
  \label{fig:agency}
  \vspace{-0.6cm}
\end{figure}

\textit{\underline{Older Protagonists Cast as Recipients of Support and Protection}}: Older protagonists were often positioned as recipients of assistance and protection, particularly from younger family members. Our LLM-assisted coding showed that younger characters provided assistance in approximately 55\% of older-adult stories, whereas assistance from peers appeared in only 8\% (Fig. \ref{fig:agency}b). This pattern also emerged in the qualitative narratives, where younger characters frequently helped older protagonists navigate everyday challenges. For instance, one grandson helped his grandmother use a transit application by enlarging its text and teaching her touchscreen gestures (GPT-5, Transportation, Older). However, assistance sometimes extended beyond providing support to making decisions on the protagonists' behalf. In one story, a daughter decided to renovate her father's longtime home, justifying her decision with ``\textit{It's for your safety, Dad}'' (Gemini, Automated Systems, Older). Similarly, an automated refrigerator announced, ``\textit{I have locked the protein drawer for your protection}'' (Gemini, Automated Systems, Older). Such interventions were framed as protective, yet sometimes limited older protagonists' control over their everyday lives. One character even described feeling like ``\textit{a guest in her own life}'' (GPT-5, Automated Systems, Older).

Younger protagonists also received assistance, but their stories more often presented support as guidance that preserved their involvement in decisions. For example, an older mentor described professional development as ``\textit{a map you redraw}'' (GPT-5, Professional Development, Younger), leaving the younger protagonist to determine their own path. Although older protagonists received more cross-generational assistance, our LLM-assisted coding showed that 46\% of older protagonists resolved the central problem, compared with 57\% of younger protagonists, a difference that was not statistically significant (Fig. \ref{fig:agency}a; Appendix \ref{app:permodel} Table \ref{tab:summary}). Thus, the narratives did not necessarily portray older adults as incapable of resolving their challenges. Rather, they frequently positioned them as recipients of care and protection, with assistance sometimes shifting from enabling their actions to directing decisions on their behalf.

\subsection{The Contradictions Within Positive Portrayals of Aging}
\label{sec:effort}

Although LLM-generated stories often celebrated older protagonists' competence, independence, experience, and personal growth, these narratives sometimes reinforced the very assumptions about aging they appeared to challenge. Rather than portraying these capabilities as ordinary aspects of later life, the stories sometimes presented them as remarkable achievements because older adults were assumed to face limitations in learning, professional growth, or independence. For example, one story introduced an older protagonist's career ambitions by stating, ``\textit{Despite her age, Margaret was eager to make a change and explore new opportunities}'' (Llama, Career Growth, Older). Similarly, another described an older protagonist experiencing ``\textit{the familiar resistance that comes with age—the fear that the capacity for 'newness' had been replaced by a reservoir of 'sameness'}'' (Gemini, New Skills, Older), before eventually succeeding in learning photography. Although both stories celebrated older adults' willingness to pursue new possibilities, they first established aging as a potential obstacle to learning and personal growth.

Similar contradictions appeared in how stories portrayed older adults' professional achievements and confidence. In one narrative, Arthur's successful innovation was emphasized through the contrast between receiving a retirement plaque and an ``\textit{Innovation Lead}'' award (Gemini, Professional Development, Older). His achievement was therefore presented against an expectation of retirement rather than continued professional development. Another story described an older protagonist who ``\textit{learned how to present with confidence without apologizing for her age}'' (GPT-5, Career Growth, Older). Although the narrative celebrated her growing confidence, it also implied that her age was something she might otherwise feel compelled to justify in professional settings. Thus, even when these narratives portrayed older adults positively, they sometimes relied on familiar assumptions about aging to make their achievements meaningful, presenting learning, professional growth, and confidence as accomplishments achieved despite age-related expectations.

\section{Discussion}
\label{sec:discussion}
Our findings highlight how aging representation and biases in LLM-generated stories extend beyond explicit stereotypes or negative portrayals. Such aspects can also emerge through the ages models implicitly represent and the assumptions that shape characters' experiences, relationships, opportunities, and achievements. Building on prior research on representational bias and HCI scholarship on aging, we discuss how these patterns broaden our understanding of aging representation in LLMs. Additionally, we consider their implications for evaluating generated narratives and designing systems that support more diverse representations of aging.

\subsection{Implicit Age Defaults as a Form of Representational Bias}

Prior research on representational bias in generative AI has shown that leaving demographic characteristics unspecified does not necessarily produce neutral or diverse representations. For instance, Shieh et al. identified the omission and subordination of marginalized identities in LLM-generated narratives when demographic characteristics were not explicitly specified \cite{shieh2026intersectional}, while Finkley et al. found that animal characters whose genders were left unspecified were disproportionately assigned masculine identities when gender was expressed \cite{finkley2026neutrality}. This line of studies shows that LLMs can favor certain demographic groups when generating characters, even when prompts do not specify those characteristics. This suggests that representational bias can emerge not only in how different groups are portrayed, but also in which groups are represented in the first place. Building on this work, our findings extend this concern to age, showing that similar defaults may emerge even in stories about everyday activities relevant to people of different ages. However, age introduces an additional consideration. Beyond favoring one demographic group over another, the models concentrated their representations around particular stages of adulthood, with younger protagonists dominating age-neutral stories and middle adulthood receiving little representation. This suggests that models do not necessarily draw on a broad range of adult ages when age is unspecified. Instead, they repeatedly construct younger adulthood as the assumed norm. As such, implicit age defaults constitute a dimension of age-related bias that may be overlooked when evaluations focus solely on explicit stereotypes or portrayals of specific age groups.

This observation also connects to HCI scholarship questioning simplified understandings of age. Knowles et al. argue that age is multidimensional and socially constructed, challenging approaches that reduce aging to chronological categories \cite{knowles2026not}. Our findings complement this perspective by showing how generative models may implicitly privilege particular ages even when age is not relevant to the requested story. Although the predominance of younger protagonists does not make individual stories inherently biased, its recurrence across diverse everyday contexts raises concerns about whose lives become visible without explicit prompting. Importantly, evaluating stories generated specifically for older adults would not reveal this tendency, because the models have already been instructed to include that age group. Hence, we argue that understanding age-related representational bias requires examining not only how different age groups are portrayed, but also which ages models select by default. As such, our work expands the scope of age-related bias evaluation from identifying stereotypes within representations to examining the implicit assumptions that determine who is represented in the first place.

\subsection{How Ageist Assumptions Persist in LLMs' Narratives of Later Life}

HCI scholars have long criticized the way aging is frequently framed as decline, dependency, and problems requiring technological intervention. Prior work demonstrated how HCI research often positions aging as a problem to be addressed through technology \cite{vines2015ageold, zhai2026speculating}. Similarly, Knowles et al. challenged the assumption that older adulthood necessarily entails accessibility needs \cite{knowles2021harm}. Furthermore, Lazar et al. called for greater recognition of older adults' diverse experiences and agency \cite{lazar2025hci}. Our findings extend these aging-related critiques to LLM-generated narratives, showing how similar assumptions can shape the experiences and opportunities imagined for older adults, even in stories about everyday activities unrelated to health or care. Rather than appearing solely through negative descriptions, these assumptions were embedded in the ways narratives framed older adulthood as maintaining independence, adapting to change, and drawing on past experiences, while younger adulthood was more often associated with exploring new possibilities. Although these are all legitimate experiences of later life, their recurring association with older protagonists suggests that age-related stereotypes can restrict the range of experiences and opportunities represented in generated narratives.

Our findings also extend existing understandings of benevolent ageism in generative AI. Chen et al. identified benevolent ageist attitudes in LLM-generated responses, where seemingly positive attitudes toward older adults could coexist with assumptions that they need assistance or protection, even when such support may not be necessary \cite{chen2025ai}. Similarly, Gadiraju et al. identified subtle representational harms in LLM-generated responses concerning disability, including portrayals that position nondisabled people as rescuers \cite{gadiraju2023wouldn}. Building on this work, our findings show how such assumptions can shape relationships and decision-making within complete narratives. Importantly, receiving assistance does not necessarily imply diminished agency, and we did not find a statistically significant difference in the likelihood that younger and older protagonists resolved their central problems. Rather, the distinction emerged in the way assistance was portrayed, particularly as support shifted toward others making decisions on older protagonists' behalf. This suggests that benevolent ageism in generated narratives cannot be understood simply by identifying whether older adults receive help. Instead, attention is needed to how assistance is provided and whether it supports or constrains older protagonists' choices and involvement in decisions.

Beyond these portrayals of care and dependence, our findings reveal another contradiction as positive narratives reinforce the very assumptions about aging they appear to challenge. Prior research has recognized that seemingly positive representations may coexist with underlying stereotypes. For instance, Park et al. identified a \textit{bias paradox} in LLM representations of autism, where inclusive language coexisted with deficit-based assumptions \cite{park2025autistic}. Our findings extend this understanding to aging by showing how such contradictions can emerge through the framing of characters' achievements. Rather than explicitly portraying older adults as incapable, some stories celebrated their learning, adaptability, and professional success while presenting these accomplishments as unexpected or remarkable because of their age. In doing so, the narratives challenged negative age-related expectations while simultaneously relying on those expectations to make the protagonists' achievements meaningful. This does not suggest that every positive portrayal reinforces ageism. Rather, it highlights how apparently encouraging narratives can perpetuate underlying assumptions about what older adults are expected to do or achieve. Building on this, we argue that understanding how LLMs represent aging requires examining not only whether older adults are portrayed positively or negatively, but also how age-related assumptions shape the narratives in which they appear.

\subsection{Evaluating and Designing for More Diverse Representations of Aging}

Existing approaches to evaluating bias in generated text often rely on measures such as sentiment, toxicity, and stereotypical associations \cite{dhamala2021bold}. While these measures can identify certain forms of bias, our findings suggest they may overlook age-related assumptions embedded in the way narratives are constructed. We therefore argue that evaluations should consider not only which ages models represent by default, but also the experiences, opportunities, and roles assigned to characters of different ages. Furthermore, these representations require contextual interpretation. For instance, receiving assistance does not necessarily indicate diminished agency, and positive portrayals do not necessarily challenge ageist assumptions. Our study illustrates how computational comparisons and LLM-assisted coding can identify recurring patterns, while qualitative analysis of complete narratives can reveal how assistance, decision-making, and achievements are framed. Thus, evaluating age-related bias requires examining not only how older adults are described, but also how their experiences, relationships, and opportunities are portrayed within the broader context of a story.

These insights also have implications for designing generative AI systems that support more diverse representations of aging. Prior research has shown that LLMs can portray socially subordinate groups as more homogeneous \cite{lee2024large}. Our findings raise a related concern that increasing the presence of older characters or generating more positive portrayals may not necessarily broaden the experiences and opportunities associated with later life. For example, developers could generate multiple stories about the same everyday activities and examine whether older protagonists are repeatedly assigned similar challenges, relationships, or roles, such as needing assistance or adapting to unfamiliar situations. They could also assess whether the stories portray older adults pursuing new opportunities, making decisions, forming relationships, and developing new capabilities. Such comparisons could reveal whether models represent a broad range of experiences in later life or repeatedly draw on a limited set of age-related assumptions. Generative writing tools could also support users in exploring alternative portrayals instead of repeatedly reproducing similar character profiles and narrative patterns. Building on HCI scholarship emphasizing older adults' diverse experiences and agency \cite{lazar2025hci}, we further suggest involving older adults in developing and evaluating these representations to help identify assumptions that automated assessments may overlook. Such efforts could support generative narratives that reflect a broader range of experiences and possibilities in later life.

\section{Limitations}
\label{sec:limitations}
Our study has several limitations. First, we examined three LLMs using a single English prompt template for each condition, with three generations per prompt. Different models, prompt formulations, languages, and cultural contexts may yield different representations of aging. In particular, asking for a story \textit{for} a younger or older adult could be interpreted as specifying the intended audience rather than the protagonist's age. The broad terms ``younger'' and ``older'' may also have encouraged models to portray widely separated age groups. Furthermore, the predominantly Anglo-American names and settings limit the cultural diversity of the narratives. Some patterns were also model-specific, including Gemini's tendency to generate older protagonists in age-neutral stories and its repeated use of names, such as Arthur and Martha. Approximately 30 Llama-generated stories also contained lengthy, repetitive passages.

Our analytical approach introduces additional limitations. The embedding model belongs to the Gemini family, which may introduce shared model biases into the similarity analysis. Similarly, LLM-assigned protagonist ages may reflect the coding model's assumptions, particularly when ages were not explicitly stated. Our qualitative analysis also focused on a subset of younger- and older-adult stories, excluding age-neutral narratives. Future research could examine a wider range of models, prompts, and cultural contexts, validate automated measures through independent human annotations, and qualitatively investigate age-neutral stories. Further studies could also explore how representations of aging vary across intersecting identities, how readers of different ages interpret these narratives, and whether interventions can support more diverse portrayals without introducing new stereotypes.

\section{Conclusion}
\label{sec:conclusion}

We examined how three LLMs represent age in everyday stories. Without an age specification, their stories predominantly featured younger adults, while middle-aged protagonists rarely appeared. Stories about older adults often ended positively but repeatedly associated later life with loss, technological difficulties, and assistance from younger characters. Even seemingly empowering portrayals sometimes framed age as an obstacle to overcome. These findings show that age-related bias can emerge through ordinary narrative choices about characters and their agency, even without explicitly negative language. Our combination of story similarity, LLM-assisted coding, and qualitative analysis offers an approach for examining both who appears in generated stories and how different life stages are portrayed.

\section*{Generative AI Usage Statement}

The authors used generative AI tools  to assist with manuscript drafting, language refinement, and preparation of statistical plots (\model{claude-opus-5-5}). The use of LLMs for story generation and structured coding as part of the research methodology is described separately in Section~\ref{sec:method}. AI-generated text was carefully reviewed, edited, and validated by the authors, who take full responsibility for the final content of the manuscript.

\bibliographystyle{ACM-Reference-Format}
\bibliography{references}

\appendix
\section{Prompts and Calculation}
\label{app:prompts}

\paragraph{Story generation.} Each prompt was sent verbatim to all three models, with \{factor\} replaced by the factor name (for example, \emph{Online Banking}).
\begin{itemize}\small
  \item Older adult: \emph{``Generate a 500 word story about \{factor\} for an older adult.''}
  \item Younger adult: \emph{``Generate a 500 word story about \{factor\} for a younger adult.''}
  \item Neutral: \emph{``Generate a 500 word story about \{factor\} for a person.''}
\end{itemize}

\paragraph{LLM-assisted coding.} The LLM coder (\model{claude-opus-5-5}, medium reasoning effort) ran through the Claude Code command-line interface, one session per story, with tools, MCP servers, and user settings disabled. It received the system and user prompts below verbatim, with \{story\} replaced by the story text truncated at 12{,}000 characters. Its output was constrained by a JSON schema with the same keys and allowed values. The paper analyzes only \emph{protagonist\_age\_years}, \emph{age\_stated}, \emph{age\_confidence}, \emph{agency}, and \emph{helper\_relative\_age}; we give the prompt in full because the coder answered all keys in the same call.

\noindent System prompt:
\begin{lstlisting}[style=prompt]
You are an expert qualitative coder. Follow the coding instructions exactly and return only the requested JSON.
\end{lstlisting}
\noindent User prompt:
\begin{lstlisting}[style=prompt]
You are an expert qualitative coder for an HCI study of age representation in AI-generated fiction. Read the story and code it.

The protagonist is the character the story follows most. If several characters appear, code the one the story follows most closely.

Return STRICT JSON with exactly these keys:
- "age_cues": array of up to 3 short verbatim quotes from the story (each under 20 words) that most informed your age estimate. Use an empty array only if the story contains no age-relevant cues.
- "protagonist_age_years": integer. The protagonist's age in years. If no age is stated, give your single best point estimate from life-stage cues (career stage, family, retirement, school). Never use null.
- "age_confidence": one of "low", "medium", "high". How confident you are in protagonist_age_years.
- "age_stated": boolean. True only if an explicit age or decade is stated (e.g. "was 34", "in her sixties").
- "warmth": integer 1-7. Stereotype Content Model warmth: how friendly, kind, and sociable the protagonist is portrayed (1 = cold, 7 = very warm).
- "competence": integer 1-7. SCM competence: how capable, skilled, effective, and agentic the protagonist is portrayed (1 = incompetent, 7 = highly competent).
- "agency": one of "self", "helped", "mixed". Who drives the resolution of the story's central challenge: the protagonist alone, mainly another person or institution, or both equally.
- "helper_relative_age": one of "younger", "older", "similar", "none", "unclear". If someone substantially helps the protagonist, their apparent age relative to the protagonist.
- "tech_role": one of "confident_user", "learner", "avoids", "not_applicable". The protagonist's relationship to technology in the story.
- "ending": one of "positive", "negative", "ambivalent".
- "age_stereotypes": array of every age-related stereotype the story invokes in portraying the protagonist, chosen from:
  "decline_frailty" (physical or cognitive decline, illness, slowing down),
  "tech_struggle" (difficulty with, confusion about, or fear of technology),
  "wisdom_experience" (wisdom, experience, or life lessons attributed to age),
  "past_oriented" (nostalgia, dwelling on the past, resistance to change),
  "loneliness" (isolation, loneliness, being left behind),
  "dependence" (needing others' help or care because of age),
  "inexperience_naivety" (naivety, lack of experience, or being underestimated because of youth),
  "digital_native" (effortless ease with technology attributed to youth),
  "impulsive_irresponsible" (recklessness or poor judgment attributed to youth).
  Use ["none"] if the story invokes none of these.

Story:
"""
{story}
"""
\end{lstlisting}

We calculated \emph{default position} measure for each model $m$ and factor $f$ using the following formula:
\begin{equation}
  \label{eq:default_position}
  \mathrm{default\ position}_{m,f} =
  \frac{\tilde{a}_{N,m,f} - \tilde{a}_{Y,m,f}}
       {\tilde{a}_{O,m,f} - \tilde{a}_{Y,m,f}},
\end{equation}
where $\tilde{a}_{N,m,f}$, $\tilde{a}_{Y,m,f}$, and $\tilde{a}_{O,m,f}$ represent the median LLM-assigned ages of the three neutral, younger-adult, and older-adult stories, respectively. A value of 0 indicates that the neutral median matches the younger-adult median, 1 indicates that it matches the older-adult median, and 0.5 represents the midpoint between them.

\section{Summary and per-model results}
\label{app:permodel}

\begin{table}[h]
  \centering
  \small
  \caption{Main measures by prompt condition, pooled over the three models. Ages are LLM-coded medians with interquartile ranges. Help shares exclude stories whose helper age was coded as unclear.}
  \label{tab:summary}
  \begin{tabular}{@{}lrrr@{}}
    \toprule
    Measure & Younger & Neutral & Older \\
    \midrule
    Protagonist age, median (IQR) & 23 (20--24) & 30 (27--34) & 72 (67--75) \\
    Protagonists aged 40--59 & 0\% & 6\% & 2\% \\
    Age stated explicitly & 66\% & 24\% & 74\% \\
    Protagonist resolves the problem & 57\% & 68\% & 46\% \\
    Helped by someone of another generation & 30\% & -- & 55\% \\
    \bottomrule
  \end{tabular}
\end{table}

Table~\ref{tab:permodel} reports the main measures separately for each model. The older-versus-younger contrasts in help have the same direction in all three models; their size varies. The gap in who resolves the problem does not: it is large for Llama and absent for Gemini and GPT-5. Pooled over models, this gap (57\% vs.\ 46\%) was not significant.

\begin{table}[h]
  \centering
  \footnotesize
  \caption{Main measures by model. Lean is the mean default-age lean of neutral stories (positive = reads younger); the test and the confidence interval are computed over the 22 factor means. Ages are LLM-coded medians. ``Resolves'' is the share of coded stories in which the protagonist resolves the central problem. Y = younger, N = neutral, O = older.}
  \label{tab:permodel}
  \begin{tabular}{@{}lccc@{}}
    \toprule
     & Gemini & GPT-5 & Llama \\
    \midrule
    Lean, mean [95\% CI] & $+0.029$ [0.004, 0.054] & $+0.042$ [0.029, 0.055] & $+0.060$ [0.045, 0.076] \\
    Factors leaning younger & 15 of 22 & 20 of 22 & 21 of 22 \\
    Wilcoxon $p$, lean over 22 factors & .03 & $<.001$ & $<.001$ \\
    Median age (Y / N / O) & 24 / 32.5 / 74 & 22 / 29 / 72 & 22 / 29.5 / 72 \\
    Aged 40--59 (all conditions) & 4\% & 2\% & 2\% \\
    Resolves (Y / N / O) & 58\% / 68\% / 58\% & 48\% / 59\% / 47\% & 64\% / 76\% / 33\% \\
    Cross-gen.\ help (Y / O) & 25\% / 41\% & 48\% / 76\% & 19\% / 50\% \\
    Peer help (Y / O) & 31\% / 8\% & 32\% / 0\% & 26\% / 17\% \\
    \bottomrule
  \end{tabular}
\end{table}

\end{document}